%% file: EuCAP2025_template.tex
\documentclass[conference,a4paper]{IEEEtran}
\usepackage{xcolor}
\usepackage{balance}
\usepackage[nolist]{acronym}
\usepackage[T1]{fontenc}

\ifCLASSINFOpdf
  \usepackage[pdftex]{graphicx}
\else
  \usepackage[dvips]{graphicx}
\fi
\usepackage{amsmath}
\ifCLASSOPTIONcompsoc
 \usepackage[caption=false,font=normalsize,labelfont=sf,textfont=sf]{subfig}
\else
 \usepackage[caption=false,font=footnotesize]{subfig}
\fi
\begin{document}

\input{acronyms.tex}
\title{On the Use of CVRP to Diagnose Faulty Elements in Antenna Arrays}

\author{\IEEEauthorblockN{
Alejandro Antón Ruiz\IEEEauthorrefmark{1},   % 1st author, 1st affiliations
John Kvarnstrand\IEEEauthorrefmark{2},    % 2nd author, 2nd affiliations
Klas Arvidsson\IEEEauthorrefmark{2},      % 3rd author, 3rd affiliations
Andrés Alayón Glazunov\IEEEauthorrefmark{1}\IEEEauthorrefmark{3} 
}                                     % ...
%\\
\IEEEauthorblockA{\IEEEauthorrefmark{1}% 1st affiliations
University of Twente, Enschede, The Netherlands, \{a.antonruiz a.alayonglazunov\}@utwente.nl}
\IEEEauthorblockA{\IEEEauthorrefmark{2}% 2nd affiliations
Bluetest AB, Gothenburg, Sweden, name.familyname@bluetest.se}
\IEEEauthorblockA{\IEEEauthorrefmark{3}% 3rd affiliations
Linköping University, Norrköping Campus, Sweden, andres.alayon.glazunov@liu.se}
}



\maketitle

\begin{abstract}
This paper investigates the application of Constrained-View Radiated Power (CVRP) for diagnosing phased array element failures, specifically focusing on on-off element failure. CVRP, similar to Partial Radiated Power (PRP), considers a specific Field-of-View (FoV) but normalizes it by the FoV area. The study explores CVRP's effectiveness in detecting failures in a 2x8 cosine element array under beam-steering conditions, accounting for random and depointing errors, angular resolution, and pattern rotation. Results indicate that CVRP can detect on-off failures based on angular resolution and error severity, under the assumption of reduced Total Radiated Power (TRP) with element failures. Additionally, CVRP is effective with partial far-field patterns, making it suitable for near-field, indirect far-field, and far-field measurement systems without requiring phase acquisition in the latter two.
\end{abstract}

\vskip0.5\baselineskip

% \fi
% \IEEEpeerreviewmaketitle

\section{Introduction}

With the increasing use of \ac{mmWave} systems for radar, \ac{5G}, and satellite communications \cite{mmWavejust}, more array antennas are being produced. Such antennas are ideal to compensate for the increased pathloss at \ac{mmWave} and, due to their directive nature, it is necessary that they can be focused where it is needed. Phased arrays enable this feature by applying beamsteering through phase, amplitude, or both, excitation of their elements.

Testing of phased arrays generally belongs to \ac{OTA} testing, since their antenna feed ports cannot generally be accessed \cite{5G_Testing_Survey}. Testing is an essential step in product development and production. In particular, \ac{OTA} testing provides a performance evaluation of the whole wireless device, phased array in this case in a controlled environment, which offers comparability between, e.g. multiple production units.

One relevant outcome of \ac{OTA} testing is the \ac{EIRP} radiation pattern. It can be obtained either from direct or indirect far-field systems, as well as near-field ones. \ac{OTA} testing has an added layer of complexity for near-field systems since phase acquisition is needed and phase synchronization of the measurement instrument with the \ac{DUT} is generally not possible, so it is needed to resort to phase recovery techniques \cite{NF2FFPhaseless}. 

A \ac{FoM} is commonly used to condense information acquired from measurements. In the case of the \ac{EIRP} radiation pattern, there are several \acp{FoM} of interest to characterize the performance of the \ac{DUT}. They can be point quantities, such as maximum \ac{EIRP}, integral quantities, such as \ac{TRP}, or partial integral quantities, such as \ac{PRP} \cite{5GAA}, and \ac{CVRP}, introduced in \cite{CVRP_orig}. The latter \acp{FoM}, by their partial integral nature, can be used as \acp{FoM} to characterize the quality of the coverage of the \ac{DUT}, i.e., how successful the \ac{DUT} is in focusing the transmitted power into a coverage area. However, it was shown in \cite{CVRP_orig} the application of \ac{CVRP} to detect faulty elements in an antenna array. One main advantage of using this partial quantity is eliminates the need to acquire the full pattern. Moreover, since \ac{CVRP} is based solely on \ac{EIRP} values, it remains applicable across any measurement range, whether in direct far-field, indirect far-field, or near-field scenarios.

In this work, we further investigate the feasibility of using \ac{CVRP} to determine if an antenna array has a certain amount of faulty elements, which was preliminary presented in \cite{CVRP_orig}, where it was applied to experimental results. There are several methods to diagnose faulty elements on an array. Some of them make use of the phase information of the radiation pattern to solve an inverse problem and retrieve the excitation coefficients \cite{INVERSE_1,INVERSE_2,INVERSE_3}. There are other methods, which include the use of machine learning, that use amplitude-only data from the far-field pattern \cite{ML_FF_1,ML_FF_2}, while others make use also of the phase information \cite{ML_NF_1}. To the best of the authors' knowledge, there is not a thorough investigation on methods to detect failed array elements using far-field amplitude-only data without the need to resort to machine learning approaches, which require a complex process to be trained and validated.

In this study, we consider a $2$ by $8$ ideal cosine elements array, and take into account the following factors: 
\begin{itemize}
    \item On-off faulty elements: elements are turned completely off. Coupling is not considered.
    \item Beamsteering
    \item Effect of pattern rotation needed when beamsteering is applied
    \item Angular resolution
    \item Systematic (depointing) errors
    \item Random (ripple) errors
\end{itemize}

\section{CVRP and its application}
\label{S2}
\ac{CVRP} was introduced in \cite{CVRP_orig}, so we just show the equation of the approximation used when we have a \ac{EIRP} radiation pattern over a discretized $\theta$ and $\varphi$ grid and the particular application used in this paper
\begin{equation}
\resizebox{.91\hsize}{!}{$CVRP\cong\frac{\Delta\varphi\Delta\theta\ }{A}\sum_{i=1}^{N-1}\sum_{j=0}^{M-1}{EIRP_{\text{Total},msk}\left(\theta_i,\varphi_j\right)\sin{\left(\theta_i\right)}},$}
\label{Eq1}
\end{equation}
where $EIRP_{Total,msk}\left(\theta_i,\varphi_j\right)$ is the sum of the \ac{EIRP} in two orthogonal polarizations in linear scale (e.g. mW), masked by the considered $[-\frac{\varphi_{FoV}}{2},\frac{\varphi_{FoV}}{2}]$ and $[0,\theta_{FoV}]$. In particular, the values of $EIRP_{Total,msk}\left(\theta_i,\varphi_j\right)$ are set to 0 for any $\theta_i,\varphi_j$ outside of the considered intervals. $A$ is the area of the (partial) sphere defined by the considered $\varphi$ and $\theta$ intervals. $\Delta\varphi$ and $\Delta\theta$ are the grid step sizes or angular resolutions, in radians, defining the number of sampling points for $\varphi$, $M=\frac{2\pi}{\Delta\varphi}$, and for $\theta$, $N=\frac{\pi}{\Delta\theta}$.

\begin{figure}
\centering
\includegraphics[width=0.6\columnwidth]{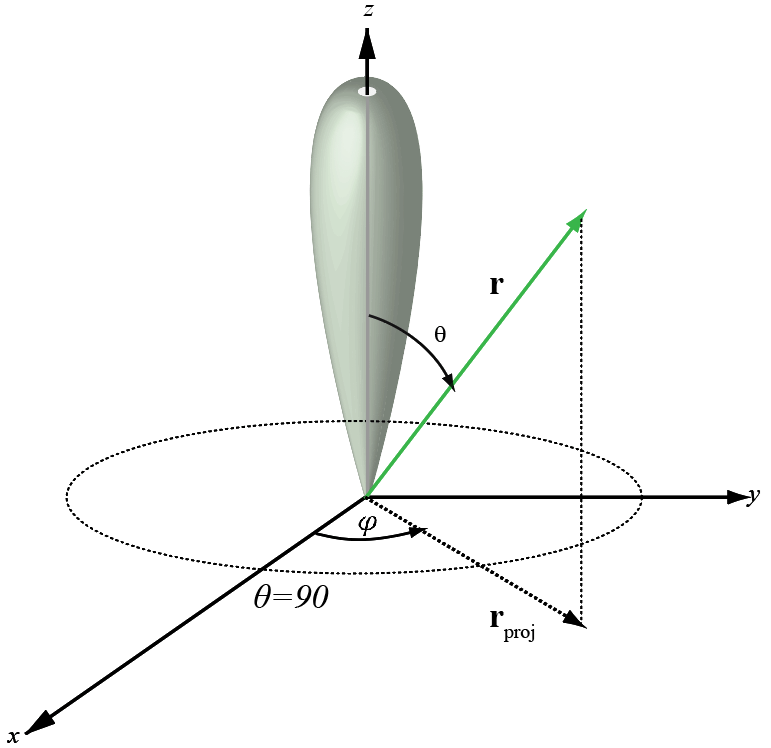}
\caption{Definition of $\varphi$ and $\theta$. Source: \cite{MATLAB_PHITHETA}.}
\label{F1}
\end{figure}

Throughout this work, we always consider a $\varphi_{FoV}$ of $360^\circ$, varying $\theta_{FoV}$. Please note that we use the definitions of $\varphi$ and $\theta$ shown in Fig.~\ref{F1}. For the limiting case where $\theta_{FoV} = 0^\circ$, we make $CVRP = EIRP_{Total}\left(0,0\right)$, which is the \ac{EIRP} corresponding to the positive z-axis. For the other limiting case, where $\theta_{FoV} = 180^\circ$, the \ac{CVRP} equals the \ac{TRP}, being $A=4\pi$, which is the area of the unit sphere.

The result of \ac{CVRP} can be interpreted as the \ac{TRP} that an ideal isotropic radiator would need to have to provide the same \ac{CVRP} over the considered area, which can be used to evaluate the power focusing in the considered area. The main difference with \ac{PRP} is the normalization by only the considered area and not the whole sphere. This enables the comparison of \ac{CVRP} values at different considered areas, defined in our case by $\theta_{FoV}$. This can be easily understood with the example of an ideal isotropic radiator, where the \ac{CVRP} would be constant over all $\theta_{FoV}$ values and equal to its \ac{TRP}, while the \ac{PRP} will be larger with an increasing considered area, until reaching the \ac{TRP} when considering the full sphere. 

One of the consequences of defining the \ac{CVRP} centered in the positive z-axis is that, whenever we have an antenna whose main lobe is not aligned with the positive z-axis, e.g. when we apply beamsteering, to perform an adequate comparison in terms of \ac{CVRP} with the broadside case we need to rotate the \ac{EIRP} radiation pattern so that the main lobe is aligned with the z-axis, which implies an interpolation of the original values. In this work, only linear interpolation of the radiation pattern \ac{EIRP} linear values is considered. 

\section{Simulation setup}
\label{S3}

\subsection{Antenna array}
\label{S31}
\begin{figure}
\centering
\includegraphics[width=0.9\columnwidth]{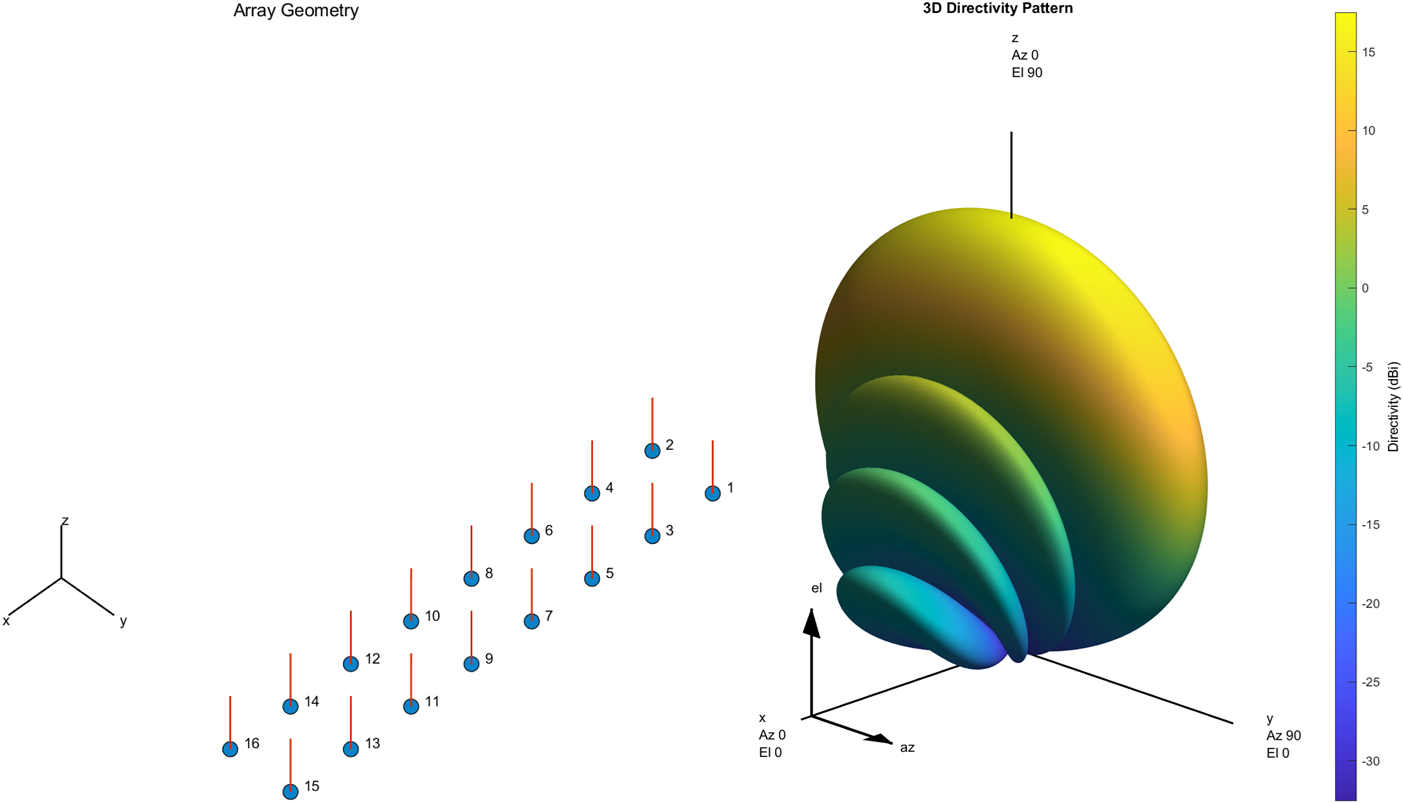}
\caption{Simulated array geometry and directivity pattern. On the left, the array geometry plot includes the indexes of each element, as well as their normals in red. Generated using the Sensor Array Analyzer app from MATLAB.}
\label{F2}
\end{figure}
The antenna array considered for this work is a $2$ by $8$ ideal cosine elements array, with $0.5\lambda$ spacing in both dimensions. The frequency of operation is $28$~GHz, but, due to the wavelength-dependent layout of the array and the idealized elements, the results are independent of the considered frequency. Its normal in its unmodified state is aligned with the z-axis.
The array layout and base radiation pattern can be observed in Fig.~\ref{F2}. The elements used are the cosine ones, located under the antenna section, with a cosine power of $[1,1]$.

\subsection{Faulty elements}
\label{S32}
This work only considers on-off element failures, which implies that the weighting of the failed element is set to $0$. Since we have the radiation pattern in terms of directivity, we convert it to \ac{EIRP} by summing a \ac{TRP} value. For similarity to \cite{CVRP_orig}, we choose the base value to be $TRP_{base}=15$~dBm. Unlike what was observed in \cite{CVRP_orig}, where deactivating array elements did not seem to impact the outputted \ac{TRP}, we consider here that the outputted \ac{TRP} is reduced proportionally to the number of failed elements. In particular, we have that
\begin{equation}
    TRP\ \left[dBm\right]=TRP_{base}+10\log_{10}{\frac{N_{el}-N_{FE}}{N_{el}}},
    \label{Eq2}
\end{equation}
where $N_el=16$ is the number of array elements, $N_{FE}$ is the number of failed elements, and 
There are $4$ considered cases in terms of element failures, denoted as ``FE X'', where ``X'' represents the failed elements. If ``X'' is $0$ it means that no elements fail. The indexes of the array elements can be observed in Fig.~\ref{F2}:
\begin{itemize}
    \item FE 0: $TRP = 15$~dBm
    \item FE 15: $TRP = 14.72$~dBm
    \item FE 7 9: $TRP = 14.42$~dBm
    \item FE 7 8 9 10: $TRP = 13.75$~dBm
\end{itemize}

\subsection{Beamsteering}
\label{S33}
\begin{figure}
\centering
\includegraphics[width=0.9\columnwidth]{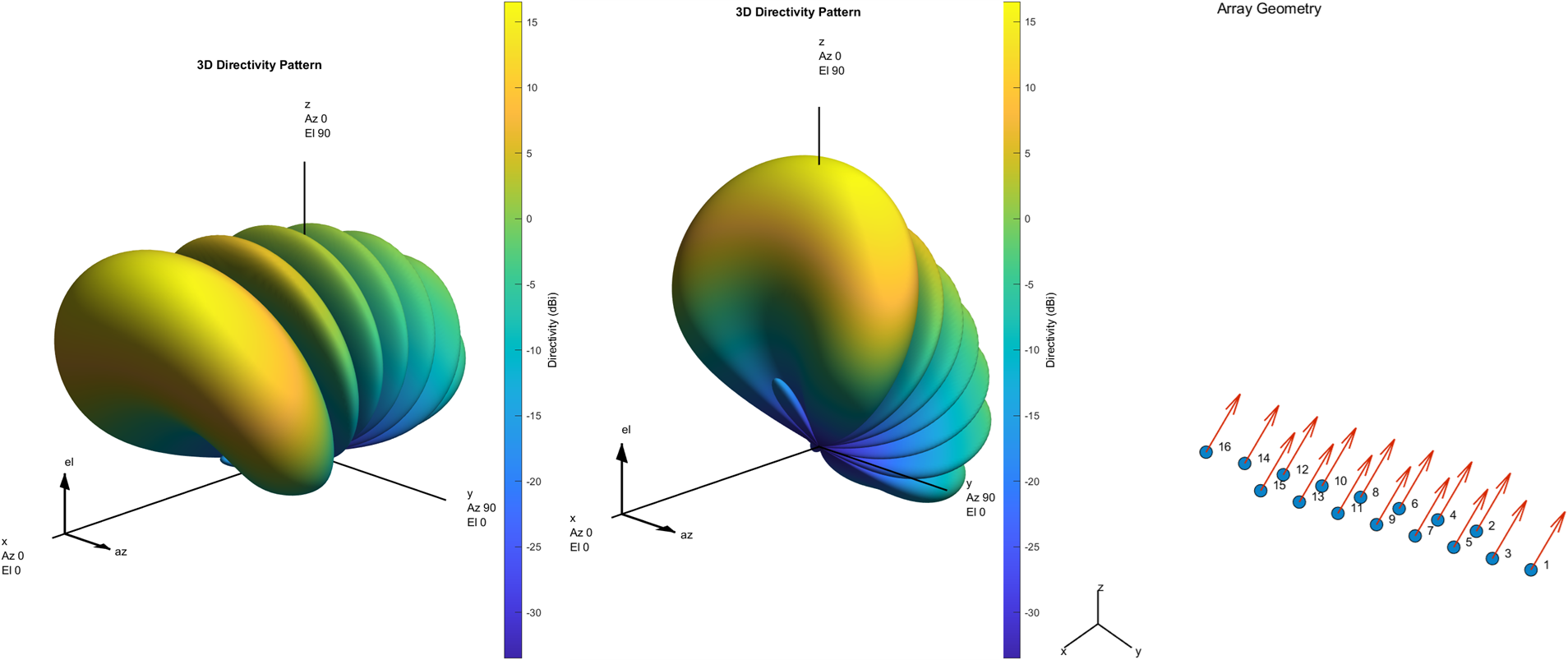}
\caption{Simulated array geometry and directivity pattern. On the left, the pattern of the array with $45^\circ$ beamsteering without rotation. In the middle, the pattern of the array with $45^\circ$ beamsteering and a $45^\circ$ clockwise rotation around the y-axis. On the right, the geometry of the array with the $45^\circ$ clockwise rotation around the y-axis. Note that the rotation is applied to both the positions of the elements and their normals. Also, note that the geometry of the array without the rotation is the same as the one depicted in Fig.~\ref{F2}.}
\label{F3}
\end{figure}
Two cases of beamsteering are considered in this work. The first is not applying any beamsteering, i.e., what can be observed in Fig.~\ref{F2}. The other case is a $45^\circ$ steering from the positive z-axis and towards the positive x-axis, contained in the xz plane, so the main beam is pointing towards $\theta=45^\circ$ and $\varphi=180^\circ$ in the coordinate system from Fig.~\ref{F1}. As stated in the end of Section~\ref{S2}, we need to rotate this radiation pattern to align the main lobe with the positive z-axis. This is done by applying a $45^\circ$ clockwise rotation around the y-axis to the radiation pattern. Without going into further detail, we just want to highlight that this implies an interpolation (linear, in this case) of the values of the original radiation pattern. Therefore, we are introducing an error. To evaluate the impact of this introduced error, we rotate the array $45^\circ$ clockwise around the y-axis, so that we have an error-free version of the pattern with the main lobe aligned with the positive z-axis to compare with the version rotated in post-processing. This is relevant since it allows us to evaluate the impact of the post-processing rotation, which will most likely be performed when taking actual measurements, as was the case in \cite{CVRP_orig}. We will refer to this rotation of the array before acquiring the radiation pattern as ``physical rotation'', in contrast with the ``post-processing rotation''. All this can be better understood with Fig.~\ref{F3}. We will denote this variable as ``SA''.

\subsection{Angular resolution}
\label{S34}
This study considers three different angular resolutions: $0.5^\circ$, $1.5^\circ$ and $5^\circ$. We consider $19$ unique values of $\theta_{FoV}$, ranging from $0^\circ$ (where the \ac{CVRP} equals the \ac{EIRP} of the positive z-axis, after any possible rotation), to $180^\circ$ (where the \ac{CVRP} equals the \ac{TRP}), with a $10^\circ$ step size. We will denote the angular resolution variable as ``RES''.

\subsection{Depointing errors}
\label{S35}
These systematic errors emulate the misalignment between the \ac{DUT}'s broadside and the measurement antenna that might occur in a real testing setup. They are applied by applying a ``physical rotation'' to the array elements and their normals. We consider depointing errors of $1^\circ$ and $3^\circ$, both of them in a clockwise fashion around the y-axis. We will denote this variable as ``DEP''.

\subsection{Random errors}
\label{S36}
We consider random errors on the acquisition of the radiation pattern. In particular, we aim to emulate a ripple error that could occur when using a \ac{CATR} system. We base our error model in \cite{CTIA120,CTIA170,CTIA173}. We consider a ripple error $\sigma_{err,dB}$ of $1$ and $2$~dB. As stated in \cite{CTIA120}, we have that $\sigma_{err,lin} = 23\cdot\sigma_{err,dB}/100$. Then, we apply
\begin{equation}
    \resizebox{.91\hsize}{!}{$EIRP_{Total,err}\left(\theta_i,\varphi_j\right)=EIRP_{Total}\left(\theta_i,\varphi_j\right)\cdot(1+\mathcal{N}(0,\sigma_{err,lin}^2)),$}
    \label{Eq3}
\end{equation}
so every point of the radiation pattern gets distorted by a random amount which is proportional to its original value, thus obtaining a distorted radiation pattern. The objective of simulating this error is to obtain a confidence interval of the \ac{CVRP} values for each $\theta_{FoV}$ and for every combination of the rest of the factors. This is attained by obtaining \ac{CVRP} values for $1000$ different distorted patterns for each $\sigma_{err,dB}$ value. Then, we estimate the standard deviation of each \ac{CVRP} value from the $1000$ acquired values ($\hat{\sigma}_{lin}$), and compute the $95\%$ \ac{CI} as
\begin{equation}
    CVRP\ \left[mW\right]=mean\left(CVRP_{sim}\right)\pm1.962\cdot{\hat{\sigma}}_{lin},
    \label{Eq4}
\end{equation}
where $1.962$ comes from the t-score for $95\%$ confidence and $999$ degrees of freedom.

\section{Results}
\label{S4}
In this section, we present the results of the simulations described in Section~\ref{S3}. We divide the results considering several factors at a time or separately. In general, we present the results as the subtraction of the result of interest and a reference value, with the objective to see if it is possible to discern between each case. Note that the ultimate goal is to find out if, considering all introduced errors, it is possible, and if so, in which cases, to distinguish the number of failed elements between the $4$ different cases.

\subsection{Effect of post-processing rotation}
\label{S41}
\begin{figure}
\centering
\includegraphics[width=0.9\columnwidth]{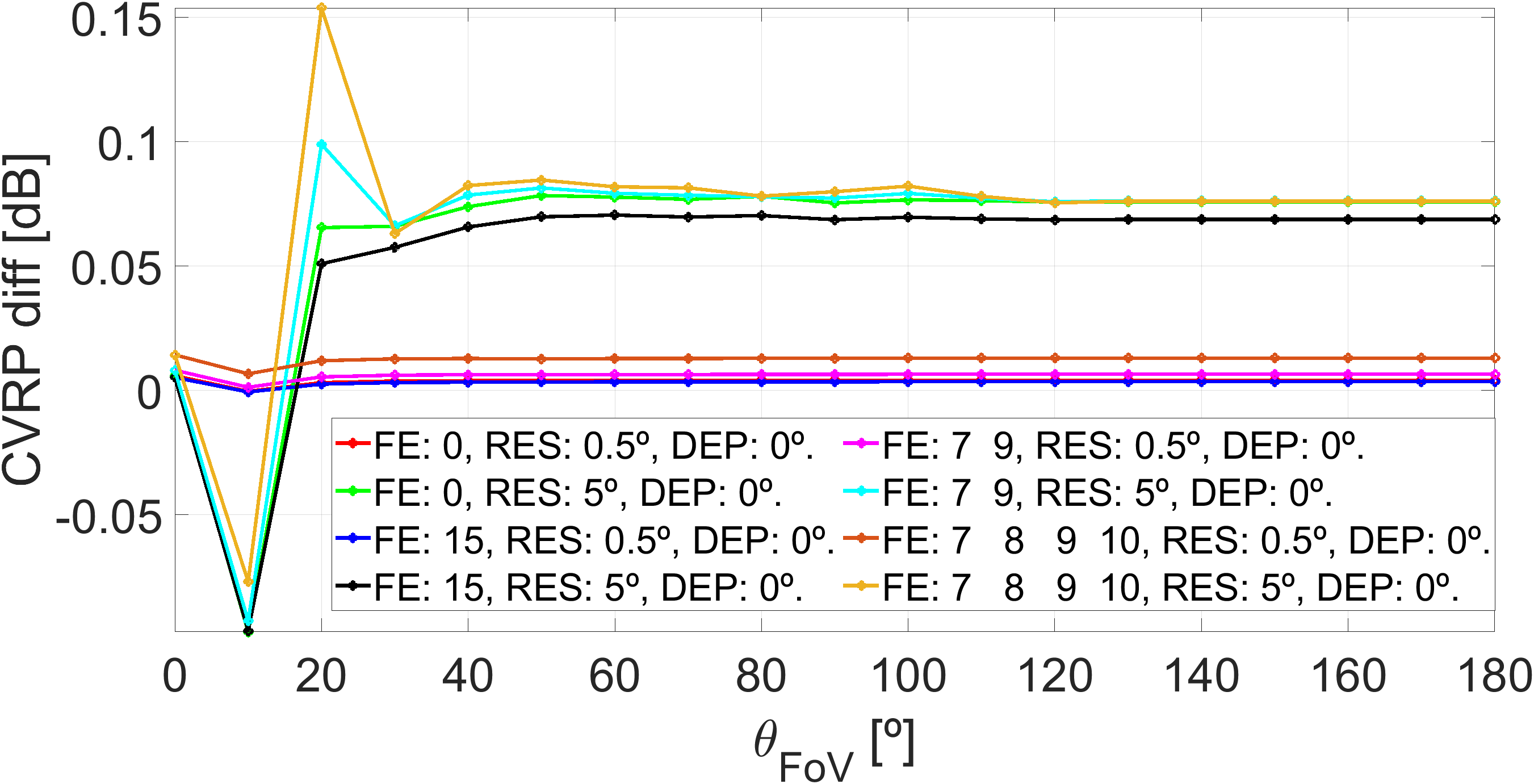}
\caption{Post-processing rotation \ac{CVRP} minus physical rotation \ac{CVRP} for $45^\circ$ beamsteering case. The same angular resolution (``RES'') is used for computing the \ac{CVRP} values to be subtracted, e.g., the case ``FE: 15, RES: 5º, DEP: 0º'' is the result of the case ``FE: 15, RES: 5º, DEP: 0º'' with post-processing rotation minus the case ``FE: 15, RES: 5º, DEP: 0º'' with physical rotation. No depointing nor random errors are considered.}
\label{F4}
\end{figure}
First, we show the impact of the post-processing rotation of the pattern in the \ac{CVRP} values by comparing them to the physical rotation ones. This quantifies the error introduced by the post-processing rotation that is needed when we perform beamsteering. In Fig.~\ref{F4}, it can be observed that, for the finest angular resolution of $0.5^\circ$, the introduced error is lower than $0.015$~dB in all cases (note that the red line is behind the dark blue line), so we can consider it negligible. For the coarsest angular resolution of $5^\circ$, the introduced error reaches a maximum of around $0.15$~dB, with a rapid oscillation for values of $\theta_{FoV}$ between $10^\circ$ and $20^\circ$, although it is drastically reduced when considering larger $\theta_{FoV}$, going lower than $0.08$~dB. We can also observe that the error is generally higher when more than one element fails, probably due to the increased irregularities of the radiation pattern, that degrade the linear interpolation accuracy. It is also worth noting that the effect for $\theta_{FoV} = 0^\circ$ is negligible in all cases.
\subsection{Effect of all errors}
\label{S42}
We have discussed the impact of the post-processing rotation in Section~\ref{S41}. However, in a realistic scenario, we would not acquire a measurement where we physically rotate the \ac{DUT}, since it might not be mechanically feasible, or, moreover, it could increase the uncertainty of the measurement due to a change in the position of the \ac{DUT} from one measurement to another.

Therefore, we now consider that we measure without any physical rotation, other than the depointing cases. We also consider that we are unaware of the depointing and that we do not correct for it by doing a post-processing rotation to revert it. This might be something obvious to do when we have all elements without failure, since it would be just a matter of searching maximum \ac{EIRP} and then rotating the pattern so that point is aligned with the positive z-axis. However, when we have failure in some elements, the maximum \ac{EIRP} is generally no longer in the intended scanning direction. 

As reference measurements, we will take the cases without any failed elements, with the corresponding angular resolution, and without any depointing. We do this because we consider that it would be a common practice to test the ``golden device'', i.e., one that we know by other means that has no failed elements with the same resolution that will be used to measure the \acp{DUT} to test and we also consider that it would be reasonable to assume that a calibration of the alignment is performed for the measurement of the ``golden device'', but not necessarily for the other \acp{DUT}.
\begin{figure}
\centering
\includegraphics[width=0.9\columnwidth]{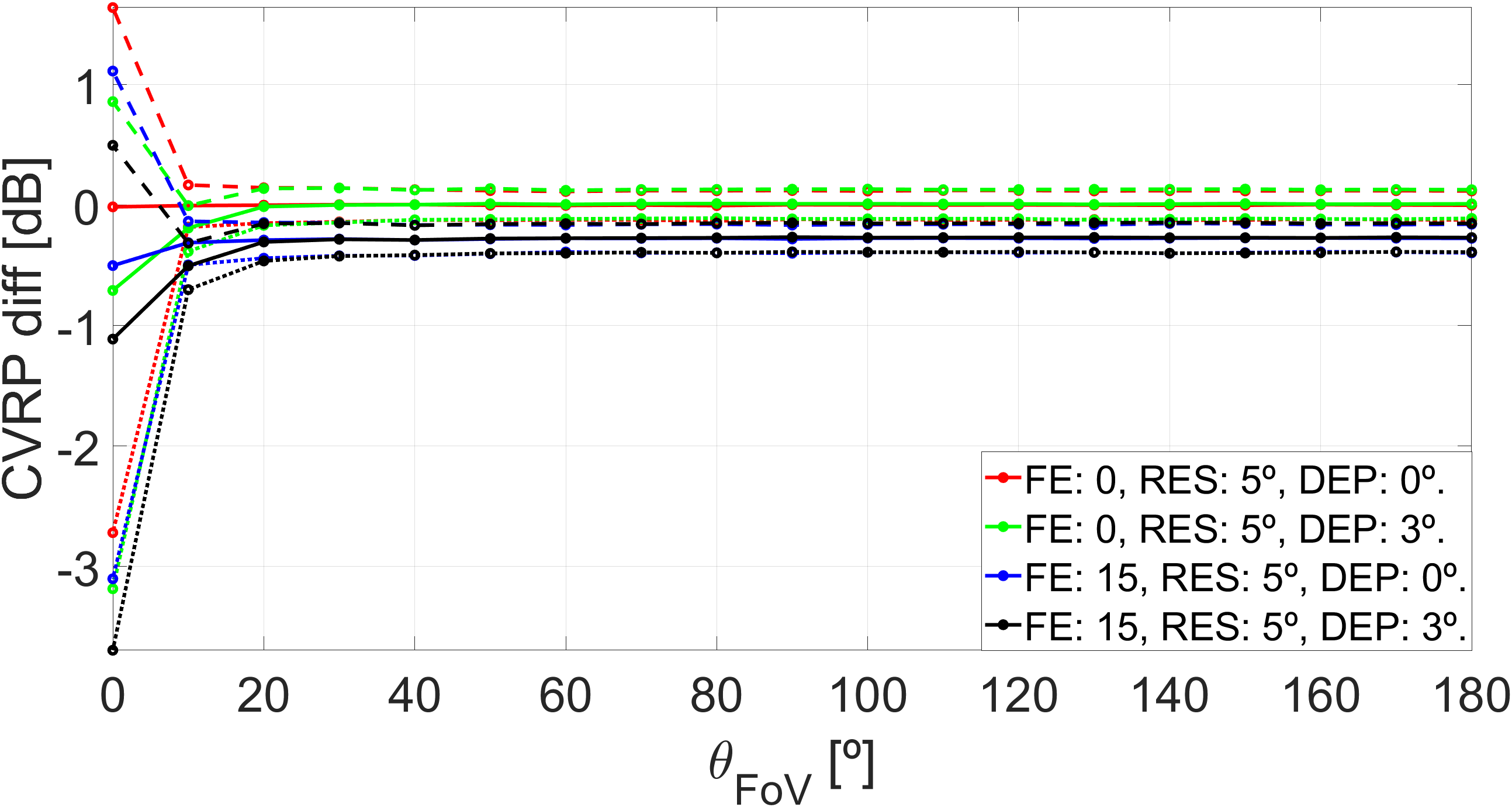}
\caption{\ac{CVRP} difference between the considered cases in the legend and the no faulty element case, with the corresponding angular resolution and without depointing, which we denote as reference. Upper bound of \ac{CVRP} \ac{CI} [dBm] minus reference \ac{CVRP} [dBm] in dashed line. Lower bound of \ac{CVRP} \ac{CI} [dBm] minus reference \ac{CVRP} in dotted line. Average of \ac{CVRP} with random errors [dBm] minus reference [dBm] in solid line. The considered $\sigma_{err,dB}$ is $1$~dB. E.g, the case ``FE: 15, RES: 5º, DEP: 3º'' is the result of the average of the $1000$ \ac{CVRP} values considering a $\sigma_{err,dB}$ of $1$~dB minus the case ``FE: 0, RES: 5º, DEP: 0º'' without any random error.}
\label{F5}
\end{figure}
\begin{figure}
\centering
\includegraphics[width=0.9\columnwidth]{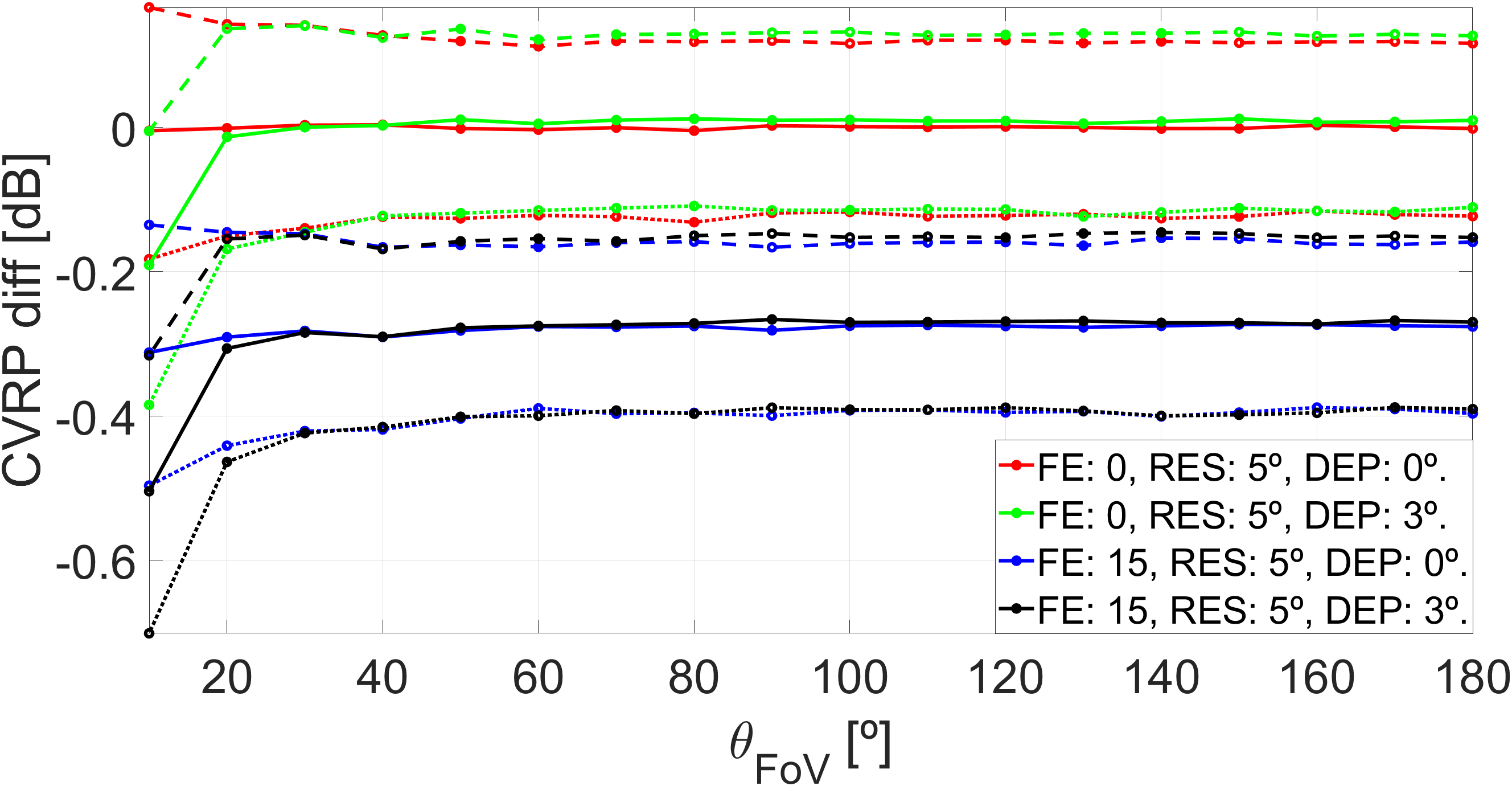}
\caption{Zoom of Fig.~\ref{F5}, covering $\theta_{FoV}$ from $10^\circ$ to $180^\circ$.}
\label{F6}
\end{figure}
We show the results that we consider that provide the most relevant information, i.e. the worst-case scenarios. 

With this into account, we can observe Fig.~\ref{F5} and Fig.~\ref{F6}, where we can see the traces for $4$ cases combining the number of faulty elements and depointing. It is worth noting that, to distinguish if we have, in this case, $1$ failed element (number $15$) or no failed elements with at least a $95\%$ of confidence, we need no overlapping between the \acp{CI}. So, we need that the lower bound of the \ac{CI} of the case without any failed elements is higher than the upper bound of the \ac{CI} of the case with the failed element. 
We represent a worst-case scenario in terms of angular resolution and depointing, but not for $\sigma_{err,dB}$. There is a clear overlap for $\theta_{FoV}\leq30^\circ$, which is rather small for $\theta_{FoV}=20^\circ$. The same holds for the $45^\circ$ beamsteered cases. If we check the other failed elements cases, we find out that it is possible to distinguish between them even at $\theta_{FoV}=10^\circ$ with a $5^\circ$ angular resolution, as long as $\sigma_{err,dB}$ is kept at $1$~dB. This holds for both the broadside and beamsteered cases. It is worth noting that, for most cases, with some exceptions when we compare $0$ and $4$ failed elements, there are overlaps for the \ac{CVRP} \acp{CI} at $\theta_{FoV}=0^\circ$. This means that, for the considered error model, $\sigma_{err,dB}$ values and the rest of the factors, it is generally not possible to infer with $95\%$ confidence if there is any failed element only based on the \ac{CVRP} at $\theta_{FoV}=0^\circ$. On the other hand, it is also relevant to note that the considered depointing does not represent a noticeable impact on the \ac{CVRP} values, not making a real difference to be able to distinguish the number of failed elements. Hence, we do not include its effect in the subsequent plots, to avoid excessive clutter. We have checked that the analysis remains virtually the same whether we consider depointing or not.
\begin{figure}
\centering
\includegraphics[width=0.9\columnwidth]{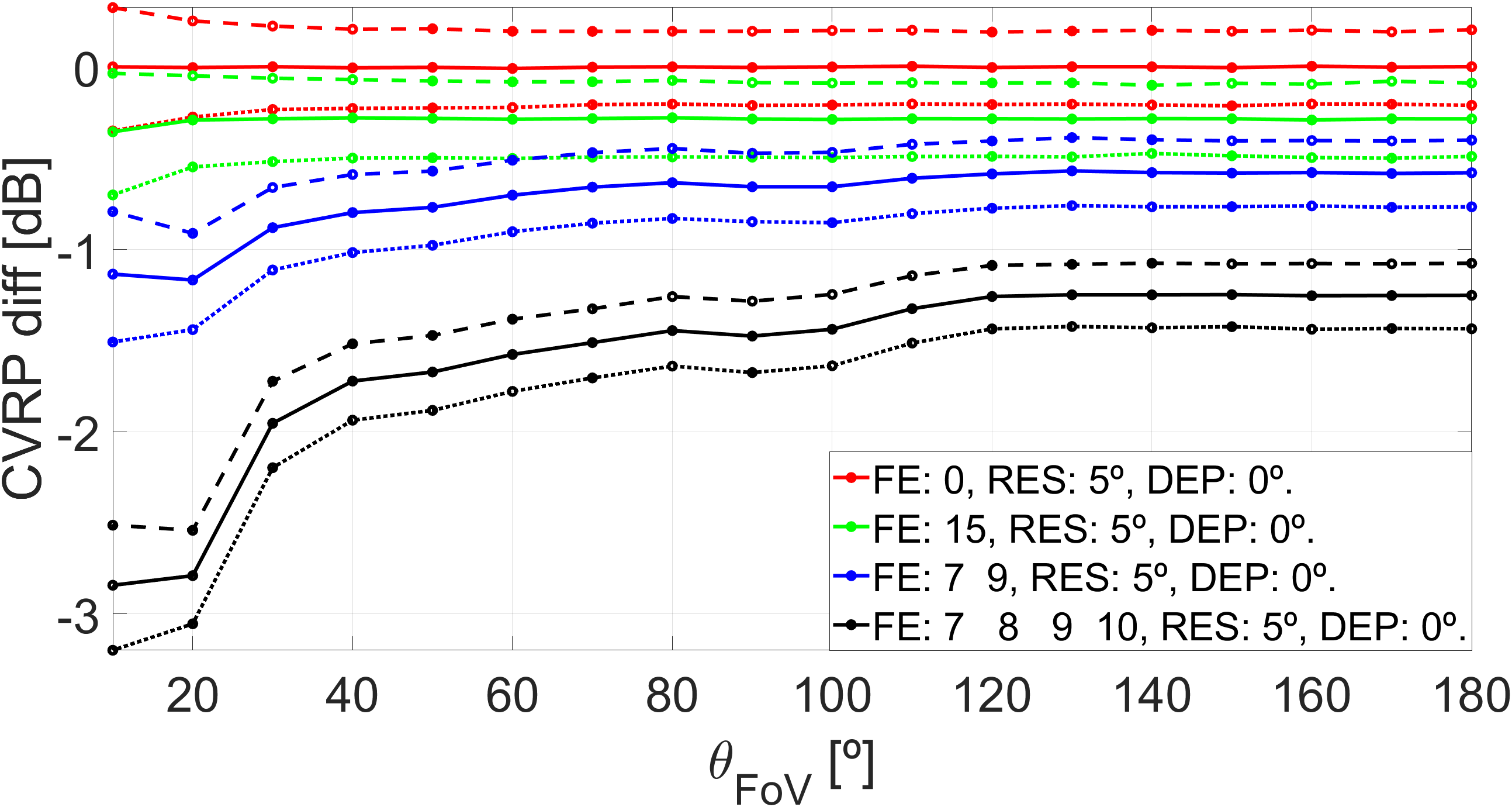}
\caption{$45^\circ$ beamsteering. $\sigma_{err,dB}=2$~dB. Angular resolution of $5^\circ$. All $4$ cases of faulty elements are considered. $\theta_{FoV}\in[10^\circ,180^\circ]$.}
\label{F7}
\end{figure}
\begin{figure}
\centering
\includegraphics[width=0.9\columnwidth]{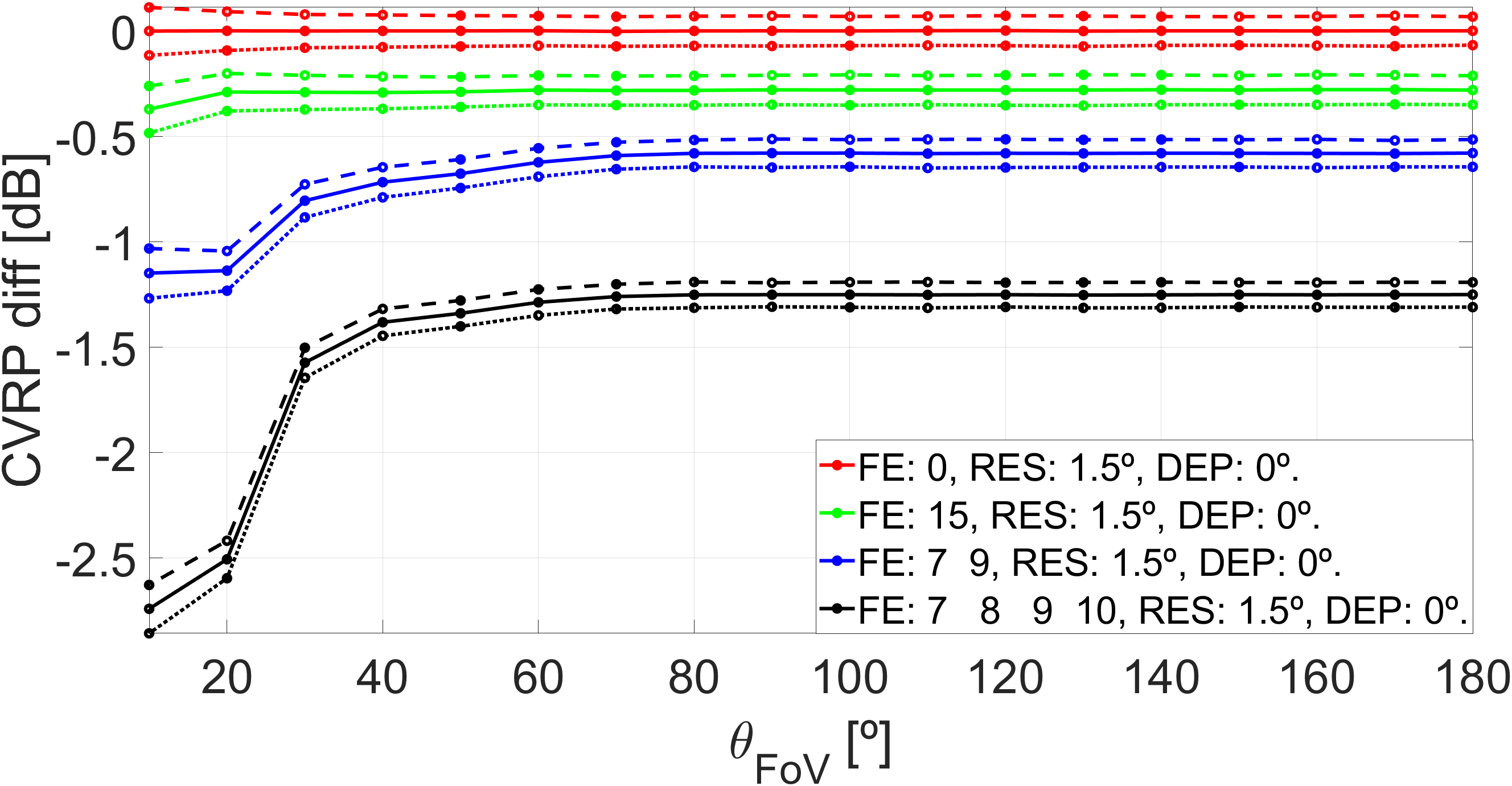}
\caption{No beamsteering. $\sigma_{err,dB}=2$~dB. Angular resolution of $1.5^\circ$. All $4$ cases of faulty elements are considered. $\theta_{FoV}\in[10^\circ,180^\circ]$.}
\label{F8}
\end{figure}
If we make $\sigma_{err,dB}=2$~dB and we keep the angular resolution at $5^\circ$ (Fig.~\ref{F7}), then we see that we are no longer able to reliably distinguish between having $0$ or $1$ failed elements, or between having $1$ or $2$ failed elements. It is still possible to reliably distinguish between having $0$ or $2$ failed elements, as well as between having $0$ or $4$, $1$ or $4$, and $2$ or $4$ failed elements, for any $\theta_{FoV}\geq10^\circ$. Finally, if we decrease the angular resolution to $1.5^\circ$, we can reliably distinguish between all $4$ faulty element cases, for any $\theta_{FoV}\geq10^\circ$. 

\balance

% An example of a floating figure using the graphicx package.
% Note that \label must occur AFTER (or within) \caption.
% For figures, \caption should occur after the \includegraphics.
% Note that IEEEtran v1.7 and later has special internal code that
% is designed to preserve the operation of \label within \caption
% even when the captionsoff option is in effect. However, because
% of issues like this, it may be the safest practice to put all your
% \label just after \caption rather than within \caption{}.
%
% Reminder: the "draftcls" or "draftclsnofoot", not "draft", class
% option should be used if it is desired that the figures are to be
% displayed while in draft mode.
%
% where an .eps filename suffix will be assumed under latex, 
% and a .pdf suffix will be assumed for pdflatex; or what has been declared
% via \DeclareGraphicsExtensions.

% Note that the IEEE typically puts floats only at the top, even when this
% results in a large percentage of a column being occupied by floats.

% Note that the IEEE does not put floats in the very first column
% - or typically anywhere on the first page for that matter. Also,
% in-text middle ("here") positioning is typically not used, but it
% is allowed and encouraged for Computer Society conferences (but
% not Computer Society journals). Most IEEE journals/conferences use
% top floats exclusively. 
% Note that, LaTeX2e, unlike IEEE journals/conferences, places
% footnotes above bottom floats. This can be corrected via the
% \fnbelowfloat command of the stfloats package.

\section{Conclusion}
\label{S5}
In this work, we considered several factors that may impact \ac{CVRP} values, and their \acp{CI}. First, we have shown that the rotation of the pattern in post-processing has a $\theta_{FoV}$-dependent behavior, which can be mitigated by improving the angular resolution (``RES''). In addition, \ac{CVRP} at $\theta_{FoV}=0^\circ$ is not useful for determining the presence or number of faulty elements, at least with the error model used. Furthermore, we showed that a $5^\circ$ angular resolution can distinguish between all four faulty element cases with at least $95\%$ confidence, as long as $\sigma_{err,dB}$ is kept at $1$~dB and for $\theta_{FoV}\geq30^\circ$. For $\sigma_{err,dB}=2$~dB, improving the angular resolution to $1.5^\circ$ allows distinction between all four cases for any $\theta_{FoV}\geq10^\circ$.

These findings imply that \ac{CVRP} is useful in diagnosing faulty elements within an array, provided the angular resolution is appropriately set. This also depends on the \ac{DUT}'s behavior when elements fail, as we assume failed elements reduce the \ac{TRP} proportionally. Importantly, full pattern acquisition is not necessary for diagnosing faulty elements using \ac{CVRP}, making this method time-efficient. Additionally, since this approach relies only on power values, phase acquisition is unnecessary unless using a near-field range, which is advantageous for \ac{OTA} testing. However, this method does not identify specific faulty elements. On the other hand, failures can be identified when the array is calibrated in the factory, but this method aims at diagnosing it at a later stage, without needing full command over the array.

Future work includes using more advanced error models, not considering a proportional decrease of \ac{TRP} with the failed elements, considering amplitude and (or) phase excitation errors instead of on-off failures, exploring different array topologies, and accounting for variations in \ac{TRP} output.


\section*{Acknowledgment}
The work of Alejandro Antón Ruiz is supported by the European Union’s Horizon 2020 Marie Skłodowska-Curie grant agreement No. 955629. Andrés Alayón Glazunov also kindly acknowledges funding from the ELLIIT strategic research environment (https://elliit.se/).

% \IEEEtriggeratref{7}
% \IEEEtriggercmd{\enlargethispage{-20cm}}


%\bibliographystyle{IEEEtran}

\bibliographystyle{IEEEtran}

\bibliography{References}

\end{document}

%% file: acronyms.tex
\begin{acronym}

\acro{2D}{Two Dimensions}%
\acro{2G}{Second Generation}%
\acro{3D}{Three Dimensions}%
\acro{3G}{Third Generation}%
\acro{3GPP}{Third Generation Partnership Project}%
\acro{3GPP2}{Third Generation Partnership Project 2}%
\acro{4G}{Fourth Generation}%
\acro{5G}{Fifth Generation}%

%%%%%A%%%%%
\acro{AI}{Artificial Intelligence}%
\acro{AoA}{Angle of Arrival}%
\acro{AoD}{Angle of Departure}%
\acro{AR}{Augmented Reality}%
\acro{AP}{Access Point}
\acro{AE}{Antenna Element}
\acro{AC}{Anechoic Chamber}
\acro{AUT}{Antenna Under Test}
%%%%%B%%%%%

\acro{BER}{Bit Error Rate}%
\acro{BPSK}{Binary Phase-Shift Keying}%
\acro{BRDF}{ Bidirectional Reflectance Distribution Function}%
\acro{BS}{Base Station}%
%%%%%C%%%%%
\acro{CA}{Carrier Aggregation}%
\acro{CDF}{Cumulative Distribution Function}%
\acro{CDM}{Code Division Multiplexing}%
\acro{CDMA}{Code Division Multiple Access}%
\acro{CPU} {Central Processing Unit}
\acro{CUDA}{Compute Unified Device Architecture}
\acro{CDF}{Cumulative Distribution Function}
\acro{CI}{Confidence Interval}
\acro{CVRP}{Constrained-View Radiated Power}
\acro{CATR}{Compact Antenna Test Range}
 
%%%%%D%%%%%
\acro{D2D}{Device-to-Device}%
\acro{DL}{Down Link}%
\acro{DS}{Delay Spread}%
\acro{DAS}{Distributed Antenna System}
\acro{DKED}{double knife-edge diffraction}
\acro{DUT}{Device Under Test}

%%%%%E%%%%%

\acro{EDGE}{Enhanced Data rates for GSM Evolution}%
\acro{EIRP}{Equivalent Isotropic Radiated Power}%
\acro{eMBB}{Enhanced Mobile Broadband}%
\acro{eNodeB}{evolved Node B}%
\acro{ETSI}{European Telecommunications Standards Institute}%
\acro{ER}{Effective Roughness}%
\acro{E-UTRA}{Evolved UMTS Terrestrial Radio Access}%
\acro{E-UTRAN}{Evolved UMTS Terrestrial Radio Access Network}%
\acro{EF}{Electric Field}
\acro{EMC}{Electromagnetic Compatibility}

%%%%%F%%%%%
\acro{FDD}{Frequency Division Duplexing}%
\acro{FDM}{Frequency Division Multiplexing}%
\acro{FDMA}{Frequency Division Multiple Access}%
\acro{FoM}{Figure of Merit}
\acro{FoV}{Field of View}
\acro{FSA}{Frequency Selective Absorber}
%%%%G%%%%%
\acro{GI}{Global Illumination} %
\acro{GIS}{Geographic Information System}%
\acro{GO}{Geometrical Optics} %
\acro{GPU}{Graphics Processing Unit}%
\acro{GPGPU}{General Purpose Graphics Processing Unit}%
\acro{GPRS}{General Packet Radio Service}%
\acro{GSM}{Global System for Mobile Communication}%
\acro{GNSS}{Global Navigation Satellite System}%
\acro{GoF}{Goodness of Fit}
%%%%%H%%%%%
\acro{H2D}{Human-to-Device}%
\acro{H2H}{Human-to-Human}%
\acro{HDRP}{High Definition Render Pipeline}
\acro{HSDPA}{High Speed Downlink Packet Access}
\acro{HSPA}{High Speed Packet Access}%
\acro{HSPA+}{High Speed Packet Access Evolution}%
\acro{HSUPA}{High Speed Uplink Packet Access}
\acro{HPBW}{Half-Power Beamwidth}
\acro{HA}{Horn Antenna}

%%%%%I%%%%%
\acro{IEEE}{Institute of Electrical and Electronic Engineers}%
\acro{InH}{Indoor Hotspot} %
\acro{IMT} {International Mobile Telecommunications}%
\acro{IMT-2000}{\ac{IMT} 2000}%
\acro{IMT-2020}{\ac{IMT} 2020}%
\acro{IMT-Advanced}{\ac{IMT} Advanced}%
\acro{IoT}{Internet of Things}%
\acro{IP}{Internet Protocol}%
\acro{ITU}{International Telecommunications Union}%
\acro{ITU-R}{\ac{ITU} Radiocommunications Sector}%
\acro{IS-95}{Interim Standard 95}%
\acro{IES}{Inter-Element Spacing}
\acro{IF}{Intermediate Frequency}

%%%%%J%%%%%

%%%%%K%%%%%
\acro{KPI}{Key Performance Indicator}%
\acro{K-S}{Kolmogorov-Smirnov}

%%%%%L%%%%%
\acro{LB} {Light Bounce}
\acro{LIM}{Light Intensity Model}%
\acro{LOS}{Line-Of-Sight}%
\acro{LTE}{Long Term Evolution}%
\acro{LTE-Advanced}{\ac{LTE} Advanced}%
\acro{LSCP}{Lean System Control Plane}%
\acro{LSI} {Light Source Intensity}

%%%%%M%%%%%
\acro{M2M}{Machine-to-Machine}%
\acro{MatSIM}{Multi Agent Transport Simulation}
\acro{METIS}{Mobile and wireless communications Enablers for Twenty-twenty Information Society}%
\acro{METIS-II}{Mobile and wireless communications Enablers for Twenty-twenty Information Society II}%
\acro{MIMO}{Mul\-ti\-ple-In\-put Mul\-ti\-ple-Out\-put}
\acro{mMIMO}{massive MIMO}%
\acro{mMTC}{massive Machine Type Communications}%
\acro{mmW}{millimeter-wave}%
\acro{MU-MIMO}{Multi-User MIMO}
\acro{MMF}{Max-Min Fairness}
\acro{MKED}{Multiple Knife-Edge Diffraction}
\acro{MF}{Matched Filter}
\acro{mmWave}{Millimeter Wave}

%%%%%N%%%%%
\acro{NFV}{Network Functions Virtualization}%
\acro{NLOS}{Non-Line-Of-Sight}%
\acro{NR}{New Radio}%
\acro{NRT}{Non Real Time}%
\acro{NYU}{New York University}%
\acro{N75PRP}{Near-75-degrees Partial Radiated Power}%
\acro{NHPRP}{Near-Horizon Partial Radiated Power}%

%%%%%O%%%%%
\acro{O2I}{Outdoor to Indoor}%
\acro{O2O}{Outdoor to Outdoor}%
\acro{OFDM}{Orthogonal Frequency Division Multiplexing}%
\acro{OFDMA}{Or\-tho\-go\-nal Fre\-quen\-cy Di\-vi\-sion Mul\-ti\-ple Access}
\acro{OtoI}{Outdoor to Indoor}%
\acro{OTA}{Over-The-Air}

%%%%%P%%%%%
\acro{PDF}{Probability Distribution Function}
\acro{PDP}{Power Delay Profile}
\acro{PHY}{Physical}%
\acro{PLE}{Path Loss Exponent}
\acro{PRP}{Partial Radiated Power}

%%%%%Q%%%%%
\acro{QAM}{Quadrature Amplitude Modulation}%
\acro{QoS}{Quality of Service}%

%%%%%R%%%%%
\acro{RCSP}{Receive Signal Code Power}
\acro{RAN}{Radio Access Network}%
\acro{RAT}{Radio Access Technology}%

%\acro{RB}{Radio Bearer}%
\acro{RAN}{Radio Access Network}%
\acro{RMa}{Rural Macro-cell}%
\acro{RMSE} {Root Mean Square Error}
\acro{RSCP}{Receive Signal Code Power}%
%\acro{RT}{Real Time}%
\acro{RT}{Ray Tracing}
\acro{RX}{receiver}
\acro{RMS}{Root Mean Square}
\acro{Random-LOS}{Random Line-Of-Sight}
\acro{RF}{Radio Frequency}
\acro{RC}{Reverberation Chamber}
\acro{RIMP}{Rich Isotropic Multipath}

%%%%%S%%%%%
\acro{SB} {Shadow Bias}
\acro{SC}{small cell}
\acro{SDN}{Software-Defined Networking}%
\acro{SGE}{Serious Game Engineering}%
\acro{SF}{Shadow Fading}%
\acro{SIMO}{Single Input Multiple Output}%
\acro{SINR}{Signal to Interference plus Noise Ratio}
\acro{SISO}{Single Input Single Output}%
\acro{SMa}{Suburban Macro-cell}%
\acro{SNR}{Signal to Noise Ratio}
\acro{SU}{Single User}%
\acro{SUMO}{Simulation of Urban Mobility}
\acro{SS} {Shadow Strength}
\acro{STD}{Standard Deviation}
\acro{SW} {Sliding Window}

%%%%%T%%%%%

\acro{TDD}{Time Division Duplexing}%
\acro{TDM}{Time Division Multiplexing}%
\acro{TD-CDMA}{Time Division Code Division Multiple Access}%
\acro{TDMA}{Time Division Multiple Access}%
\acro{TX}{transmitter}
\acro{TZ}{Test Zone}
\acro{TRP}{Total Radiated Power}

%%%%%U%%%%%

\acro{UAV}{Unmanned Aerial Vehicle}%
\acro{UE}{User Equipment}%
\acro{UI}{User Interface}
\acro{UHD}{Ultra High Definition}
\acro{UL}{Uplink}%
\acro{UMa}{Urban Macro-cell}%
\acro{UMi}{Urban Micro-cell}%
\acro{uMTC}{ultra-reliable Machine Type Communications}%
\acro{UMTS}{Universal Mobile Telecommunications System}%
\acro{UPM}{Unity Package Manager}
\acro{UTD}{Uniform Theory of Diffraction} %
\acro{UTRA}{{UMTS} Terrestrial Radio Access}%
\acro{UTRAN}{{UMTS} Terrestrial Radio Access Network}%
\acro{URLLC}{Ultra-Reliable and Low Latency Communications}%
\acro{UHRP}{Upper Hemisphere Radiated Power}%

%%%%%V%%%%%
\acro{V2V}{Vehicle-to-Vehicle}%
\acro{V2X}{Vehicle-to-Everything}%
\acro{VP}{Visualization Platform}%
\acro{VR}{Virtual Reality}%
\acro{VNA}{Vector Network Analyzer}
\acro{VIL}{Vehicle-in-the-loop}

%%%%%W%%%%%
\acro{WCDMA}{Wideband Code Division Multiple Access}%
\acro{WINNER}{Wireless World Initiative New Radio}%
\acro{WINNER+}{Wireless World Initiative New Radio +}%
\acro{WiMAX}{Worldwide Interoperability for Microwave Access}%
\acro{WRC}{World Radiocommunication Conference}%

%%%%%X%%%%%
\acro{xMBB}{extreme Mobile Broadband}%

%%%%%Z%%%%%
\acro{ZF}{Zero Forcing}

\end{acronym}